\documentclass[letterpaper, 10 pt, conference]{ieeeconf}  

\IEEEoverridecommandlockouts                              

\usepackage{graphicx} 
\usepackage{subfigure}
\usepackage{mathptmx} 
\usepackage{times} 
\usepackage{amsmath} 
\usepackage{amssymb}  
\usepackage{mathtools} 
\usepackage{enumerate} 
\usepackage{booktabs}
\usepackage{soul}
\usepackage{color}
\usepackage[svgnames]{xcolor}
\usepackage{cite}
\usepackage{ragged2e}
\usepackage{mathrsfs}
\usepackage{lipsum}

\sethlcolor{LightGrey}
\graphicspath{{figures/}}

\newcommand{\T}{^{\mathrm{T}}}							  
\newcommand{\B}[1]{\if#1\relax\bm{#1}\else\mathbf{#1}\fi} 
\newcommand{\R}[1]{\mathrm{#1}}						      

\makeatletter
\def\amsbb{\use@mathgroup \M@U \symAMSb}
\makeatother

\title{\LARGE \bf
Control of Painlev\'e Paradox in a Robotic System
}

\author{Davide Marchese$^{1}$, Marco Coraggio$^{1}$, S. John Hogan$^{2}$, and Mario di Bernardo$^{1, 2}$
\thanks{*This work was supported by (i) the Italian Ministry of Education, University and Research (Decree 976, 29/12/2014 – Art. I), in the form of grant awarded to Davide Marchese, in order to visit the University of Bristol, and (ii) by COINOR (Centro di Servizio di Ateneo per il Coordinamento di Progetti Speciali e l’Innovazione Organizzativa) and Compagnia di San Paolo, in the form of travel support and compensation attributed to John Hogan while visiting the University of Naples in June 2017.}
\thanks{$^{1}$ Department of Electrical Engineering and Information Technology, University of Naples Federico II, Via Claudio 21, 80125 Naples, Italy;
    {\tt\small davide.marchese@tim.it}, {\tt\small \{marco.coraggio, mario.dibernardo\}@unina.it}. }%
\thanks{$^{2}$ Department of Engineering Mathematics, University of Bristol, BS8 1UB, Bristol, U.K.;
    {\tt\small s.j.hogan@bristol.ac.uk}.}%
}

\begin{document}

\maketitle
\thispagestyle{empty}
\pagestyle{empty}

\begin{abstract}
The Painlev\'{e} paradox is a phenomenon that causes instability in mechanical systems subjects to unilateral constraints. 
While earlier studies were mostly focused on abstract theoretical settings, recent work confirmed the occurrence of the paradox in realistic set-ups.
In this paper, we investigate the dynamics and presence of the Painlev\'e phenomenon in a two-links robot in contact with a moving belt, through a bifurcation study.
Then, we use the results of this analysis to inform the design of control strategies able to keep the robot sliding on the belt and avoid the onset of undesired lift-off.
To this aim, through numerical simulations, we synthesise and compare a PID strategy and a hybrid force/motion control scheme, finding that the
latter is able to guarantee better performance and avoid the onset of bouncing motion due to the Painlev\'e phenomenon.
\end{abstract}

\section{Introduction}
\label{sec:introduction}

Most people have experienced at least once the annoying high-pitched sound that chalk may produce when pressed against a blackboard.
As it is now well known \cite{CHAMPNEYS2016}, the sound is the result of fast vibrations of the piece of chalk that quickly and repeatedly detaches from the blackboard and comes back into contact with it.
This phenomenon is paradoxical as the more one presses the chalk against the surface, the more likely bouncing motion becomes.
This type of oscillatory behaviour is not only annoying but can be costly and troublesome when it manifests in practical applications.
For example, the repeated lift of an automated tool performing a cut leads to imprecise processing, resulting in unusable goods or ones with reduced value \cite{Ibrahim1994}.
Moreover, in an assembly line, a robotic arm used grasping objects from a moving belt may abruptly be pushed away from the belt, resulting in decreases in speed and accuracy  \cite{Assembly2009}. 

The phenomenon described above was named after Paul Painlev\'e, who, in 1905, published the first studies related to the paradox, providing a mathematical model.
In particular, in \cite{Painleve1905}, he analysed the dynamics of a rigid stick sliding on a surface, showing that, assuming a Coulomb friction law, when the friction coefficient was higher than a certain threshold value, a non-trivial phenomenon occurs.
Namely, the solution to the differential equations describing the motion of the stick may become indeterminate or inconsistent, in the sense that the model would predict the stick to penetrate the rigid surface, which clearly is not realistic.
In the following years, many mathematicians and scientists have been interested in the study of this paradoxical phenomenon, but, as pointed out by Champneys in \cite{CHAMPNEYS2016}, to this date, all the ways in which the stick can enter the inconsistent or indeterminate solution modes have not been determined analytically.
For this reason, most of the research follows a numerical or experimental approach in the investigation of the problem. 

For example, in \cite{loststed1982}, L\"{o}tstedt created a digital simulation of the dynamics of rigid mechanical systems under unilateral constraints, in order to study the Painlev\'{e} phenomenon. 
In \cite{Or2012} and \cite{Burns2017}, numerical simulations are used to investigate how the paradox affects the motion of an inverted pendulum sliding on an inclined plane and that of a double pendulum, respectively.
Furthermore, in \cite{leine2016}, Leine et al. studied through numerical simulations the paradox in a specific two-masses system, called \textit{frictional impact oscillator}.
They showed that the critical friction value was strictly linked to the masses ratio, and the Painlev\'{e} paradox was the cause of a Hopf bifurcation, in which a sliding equilibrium loses its stability and a periodic bouncing motion appears.
Similar results can be found in \cite{LANCIONI2009}.

Another system whose motion is influenced by the paradox, is the prismatic revolute robotic set-up analysed in \cite{elkaranshawy2017solving} via numerical simulations.
It is also important to highlight that the phenomenon studied by Painlev\'{e} can influence the motion of walking robots.
As a matter of fact, most of the passive walking models such as the \emph{compass biped} or the \emph{rimless wheel} \cite{Collins1082}, assume that there is a frictional sticking contact between the foot and the surface, whereas in reality there is always a slipping of the foot.
For instance, the numerical results obtained in \cite{Or2014} showed that regular periodic gait can be subject to an instability, related to the Painlev\'{e} phenomenon. 
Moreover, in recent work by Zhao et al. \cite{zhao2008experimental}, the occurrence of the Painlev\'{e} paradox was demonstrated experimentally in a two link robotic arm whose end effector is in contact with a sliding belt. It was shown that for certain parameter values, the arm can lift off from the moving belt, showing, for the first time in the literature, a physical demonstration of the paradox in a realistic robotic set-up.

To the best of our knowledge, the only instance of a control strategy employed to avoid the onset of the paradox is described in \cite{liang2012relative}, where a PID regulator is used to control the sliding of a two-links robot on a vertical wall.

The contribution of our work is twofold.
Firstly, we extend the analysis of the system originally presented in \cite{zhao2008experimental} unfolding the bifurcation mechanisms behind the occurrence of the Painlev\'e paradox.
Secondly, we exploit this new information to synthesise appropriate control strategies to prevent the paradox from taking place; being this one of the very few attempts at using active controllers to address the problem, along with \cite{liang2012relative}. 
In particular, in order to better understand the conditions that trigger the onset of the phenomenon, we present a characterisation of the steady state dynamics for different values of the velocity of the belt.
Specifically, we find that the paradox manifests itself only when the speed of the belt is in a certain critical interval. 
However, we show that, even when the velocity of the belt is within that critical interval, some control strategies may be employed to avoid the undesired lift-off and bouncing motion stemming from it.
In particular, we show how a PID regulator and a hybrid force/motion control scheme can be exploited to reach some positioning control goals while keeping the robot in a region of the phase space such that the paradox is not triggered.
According to our simulations, the hybrid control shows the most promising results, representing an innovation with respect to \cite{liang2012relative}, where only a PID control strategy was used.
Our results nicely combine bifurcation analysis with control system design, offering a novel approach for the active suppression of the Painlev\'e paradox in realistic mechanical systems.

\section{Bifurcation analysis}
\label{sec:bifurcation analysis}

\subsection{Model description}
\label{subsec:model_description}

\begin{figure}[t]
    \centering {
        {\includegraphics[width=0.5\columnwidth]{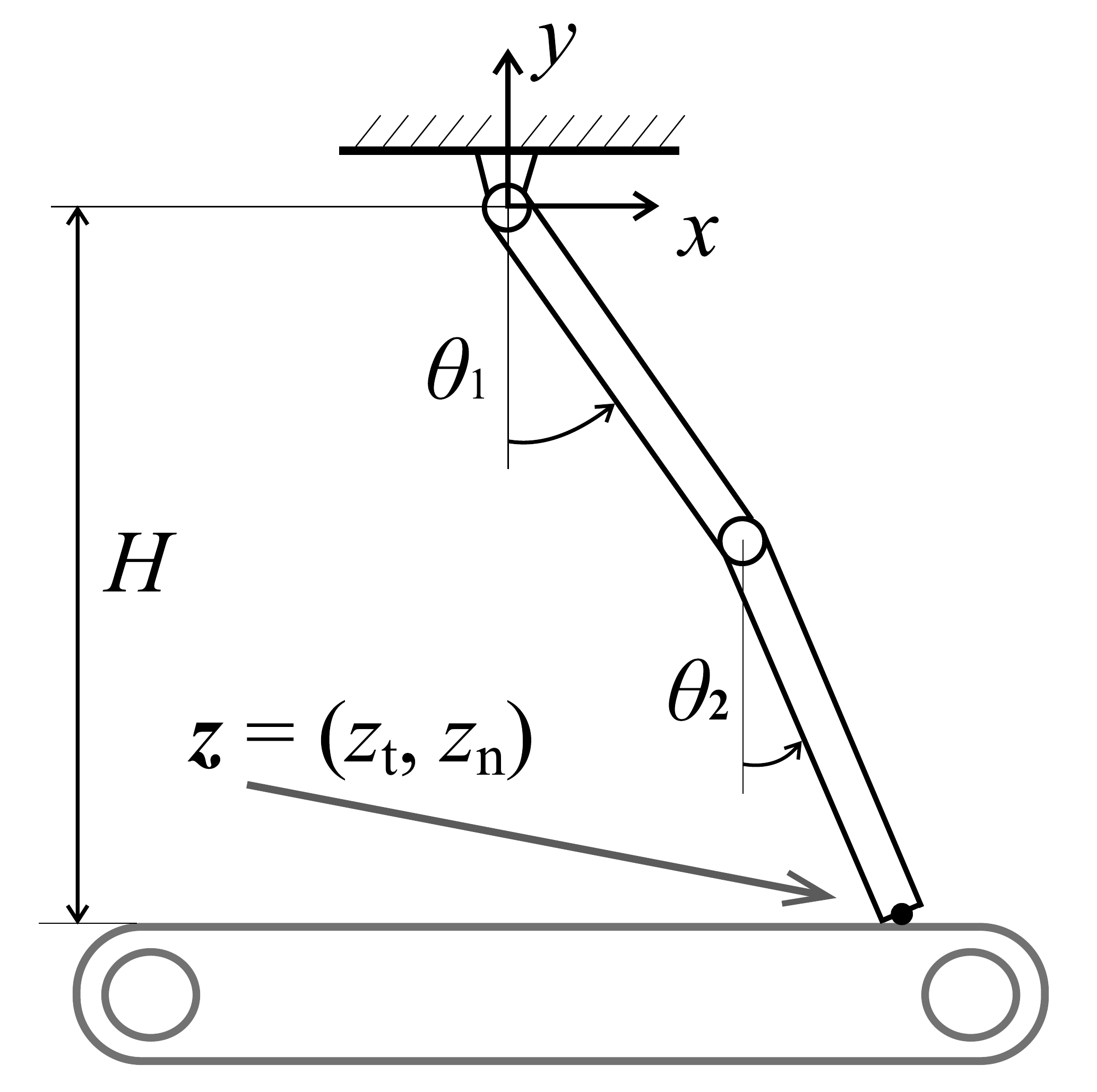}}
        \caption{A double-revolute robotic arm on a moving belt.}
        \label{fig:robot_representation}
    }
\end{figure}
We consider a two-links mechanical set-up as that represented in Figure \ref{fig:robot_representation}.
Rotational dashpots with damping coefficient $\sigma$ are present in both the joints, and a rotational spring with elastic constant $k$ is mounted in the lower joint; moreover, the belt is moving at a speed $v_\R{belt}$.
We define the generalised coordinates 
$\B{q} \triangleq \begin{bmatrix} \theta_1 & \theta_2 \end{bmatrix}\T$, 
the coordinates of the end effector
$\B{z} \triangleq \begin{bmatrix} z_\R{t} & z_\R{n} \end{bmatrix}\T$,
and, for later use, the state vector 
$\B{x} \triangleq \begin{bmatrix} \theta_1 & \dot{\theta}_1 & \theta_2 & \dot{\theta}_2 \end{bmatrix}\T$.
Then, a mathematical model of the system can be given as
\begin{equation}\label{eq:motion_equation}
\ddot{\B{q}} = \B{M}^{-1} (-\B{w} - \B{c} + \B{J}\T \B{f} + \B{u}),
\end{equation}
where:
\begin{itemize}
    \item $\B{M}$ is the mass matrix;
    \item $\B{w}$ contains the torques determined by the elastic force, viscous friction, and gravity;
    \item $\B{c}$ are the torques determined by the centrifugal and Coriolis forces;
    \item $\B{J}$ is the Jacobian, defined by the relation $\dot{\B{z}} = \B{J} \dot{\B{q}}$;
    \item $\B{u} = \begin{bmatrix} u_1-u_2 & u_2\end{bmatrix}\T$ contains the control torques, with $u_1$ and $u_2$ being the torques applied to the first and the second joint, respectively;
    \item $\B{f} = \begin{bmatrix}f_\R{t} & f_\R{n}\end{bmatrix}\T$ are the contact forces acting on the end effector, with $f_\R{n}$ being the normal reaction and $f_\R{t}$ being the Coulomb friction. 
    In particular, $f_\R{t} = - \mu \R{sign} (\dot{z}_\R{r}) f_\R{n}$, where $\dot{z}_\R{r} \triangleq \dot{z}_\R{t} - v_{\R{belt}}$ is the velocity of the contact point with respect to the belt and $\mu$ is the friction coefficient.    
\end{itemize}
The expressions of the above quantities are
\begin{equation*}
\begin{aligned}
\B{M} &= \begin{bmatrix}
\frac{4}{3} m l^{2} & \frac{ml^{2}}{2} \R{cos} (\theta_2 - \theta_1) \\[1 ex]
\frac{ml^{2}}{2} \R{cos} (\theta_2 - \theta_1) & \frac{ml^{2}}{3}
\end{bmatrix}, \\
\B{w} &= \begin{bmatrix}
\frac{3mgl}{2}\R{sin}\theta_1-k(\theta_2-\theta_1+ \alpha_0)-\sigma(\dot{\theta}_2- 2\dot{\theta}_1)\\[1 ex]
\frac{mgl}{2}\R{sin}\theta_2+k(\theta_2-\theta_1+ \alpha_0)+ \sigma(\dot{\theta}_2-\dot{\theta}_1)
\end{bmatrix}, \\
\B{c} &= \begin{bmatrix}
\frac{ml^{2}}{2}\dot{\theta}_2^2\R{sin}(\theta_{1}-\theta_{2})\\[6 pt]
\frac{ml^{2}}{2}\dot{\theta}_1^2\R{sin}(\theta_{2}-\theta_{1})
\end{bmatrix}, \\
\B{J} &= \begin{bmatrix}
\B{j}_1\T \\[1ex] \B{j}_2\T
\end{bmatrix} = 
\begin{bmatrix}
l\R{cos}\theta_1 & l\R{cos}\theta_2\\[1 ex]
l\R{sin}\theta_1 & l\R{sin}\theta_2
\end{bmatrix}.
\end{aligned}
\end{equation*}
The values of the parameters are set using the experimentally derived ones reported in \cite{zhao2008experimental} and are given in Table \ref{tab:parameter_values}.

\begin{table}[t]
    \caption{Robot parameters}
    \label{tab:parameter_values}
    \begin{center}
        \begin{tabular}{@{}llll@{}}
        	\toprule
        	Parameter            & Symbol         & Value       & Unit        \\ \midrule
        	belt speed           & $v_{\R{belt}}$ & $[-1,-0.1]$ & m/s         \\
        	friction coefficient & $\mu$          & $[0.1,1]$   & -           \\
        	links mass           & $m$            & $0.12$      & kg          \\
        	links length         & $l$            & $0.21$      & m           \\
        	damping coefficient  & $\sigma$       & $0.005$     & N$\cdot$s/m \\
        	elastic constant     & $k$            & $1.3$       & N/m         \\
        	robot height         & $H$            & $0.3775$    & m           \\
        	spring rest position & $\alpha_0$     & $13.72$     & degrees     \\ \bottomrule
        \end{tabular}
    \end{center}
\end{table}

Model \eqref{eq:motion_equation} can be recast in terms of the position $\B{z}$ of the end effector as
\begin{equation}\label{eq:motion_equation_end_effector}
\ddot{\B{z}} = -\B{J}\B{M}^{-1} (\B{w} + \B{c} - \B{u}) + \B{Q} \B{f} + \B{s},
\end{equation}
where $\B{Q} \triangleq \B{J} \B{M}^{-1} \B{J}\T = (Q_{i,j})$, $i,j = 1,2$, and $\B{s}$ is the centripetal acceleration, given by
\begin{equation*}
\B{s} = \begin{bmatrix}
s_1 \\[1 ex]
s_2
\end{bmatrix} = \begin{bmatrix}
-l(\dot\theta_1^2\R{sin}\theta_1+\dot\theta_2^2\R{sin}\theta_2) \\[1 ex]
l(\dot\theta_1^2\R{cos}\theta_1+\dot\theta_2^2\R{cos}\theta_2)
\end{bmatrix}.
\end{equation*} 
 Model \eqref{eq:motion_equation_end_effector}  can be expressed componentwise as
 \begin{equation}\label{eq:motion_equation_end_effector_t}
 \ddot{z}_\R{t} = -\B{j}_1^\R{T}\B{M}^{-1}(\B{w}+\B{c}-\B{u})
 + f_\R{n} (- \mu \R{sign} (\dot{z}_\R{r}) Q_{1,1} + Q_{1,2} ) 
 + s_1,
 \end{equation}
\begin{equation}\label{eq:motion_equation_end_effector_n}
\ddot{z}_\R{n} = -\B{j}_2^\R{T}\B{M}^{-1}(\B{w}+\B{c}-\B{u})
+ f_\R{n} (- \mu \R{sign} (\dot{z}_\R{r}) Q_{2,1} + Q_{2,2} ) 
+ s_2.
\end{equation}
For a fixed value of $\mu$, when $\dot{z}_\R{r} > 0$, we will show that there exists a region in the state space, say $\mathcal{R}^+ \subseteq \mathbb{R}^4$, such that when the state vector $\B{x} \in \mathcal{R}^+$ the paradox is triggered.
Differently, when $\dot{z}_\R{r} < 0$, the paradox  manifests itself if $\B{x} \in \mathcal{R}^- \subseteq \mathbb{R}^4$.
However, the sets of positions and velocities represented by $\mathcal{R}^+$ are symmetrical with respect to the $y$-axis to those contained in $\mathcal{R}^-$.
Therefore, for the sake of simplicity, we can limit our analysis to the case that $\dot{z}_\R{r}$ is positive; the results might then easily be extended to the case that $\dot{z}_\R{r}$ is negative by simply taking into account the symmetry between $\mathcal{R}^+$ and $\mathcal{R}^-$.
Defining the functions $p : \mathbb{R}^3 \rightarrow \mathbb{R}$ and $b : \mathbb{R}^6 \rightarrow \mathbb{R}$, given by
\begin{align}
b \triangleq &- \B{j}_2^\R{T} \B{M}^{-1} (\B{w} + \B{c} - \B{u}) + s_2, \\ 
p \triangleq &- \mu Q_{2,1}+Q_{2,2},
\end{align} 
we can rewrite \eqref{eq:motion_equation_end_effector_n} as
\begin{equation}\label{eq:normal_acceleration}
\ddot{z}_\R{n} = b (\B{q},\dot{\B{q}},\B{u}) + p(\B{q},\mu) f_\R{n}.
\end{equation}
In \eqref{eq:normal_acceleration}, the physical meaning of the newly introduced functions $b$ and $p$ is more evident.
$b$ is the \emph{free normal acceleration}, i.e. the normal acceleration in the absence of contact forces, whereas $p$ determines how the normal reaction $f_\R{n}$ influences the normal acceleration $\ddot{z}_\R{n}$ of the end effector. 

\begin{figure}[t]
    \centering {
        {\includegraphics{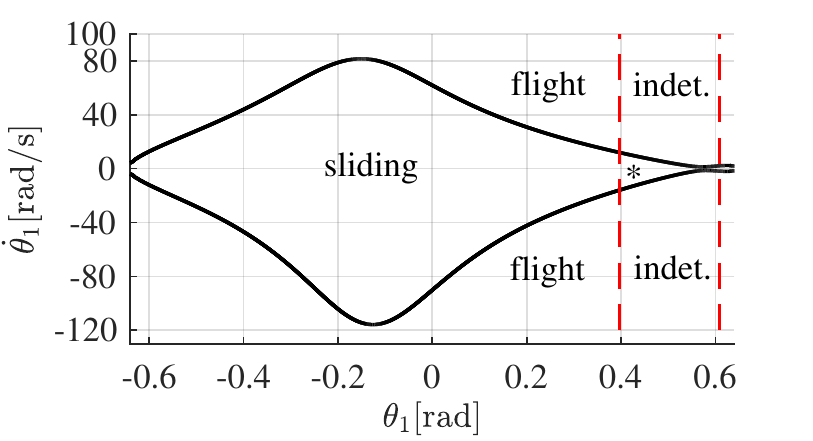}}
        \caption{Different modes of solution for different values of $\theta_1$ and $\dot{\theta}_1$. Here $\mu = 0.6$, $\B{u} = \B{0}$, and $\theta_2$ and $\dot{\theta}_2$ are chosen in order to have the tip of the robot in contact with the belt. ``indet.'' stands for indeterminate and ``$*$'' stands for inconsistent.
            The black solid line is the place where $b = 0$, while the dashed red lines are the places where $p = 0$.}
        \label{fig:phase_plane_open_loop}
    }
\end{figure}

When $z_\R{n} = -H$, the end effector is in contact with the moving belt, reproducing a situation analogous to that originally investigated by Painlev\'e.
As explained in \cite{Painleve1905} for a more general case, if $\mu$ is greater than a critical value $\mu_\R{c}$, system \eqref{eq:motion_equation_end_effector} can display four different types of solution, depending on the signs of $b$ and $p$, which in turn depend on $\B{q}$, $\dot{\B{q}}$, $\B{u}$, and $\mu$.
The first two modes, \emph{sliding} and \emph{flight}, are associated to solutions to the motion equation \eqref{eq:normal_acceleration}, both characterized by $p > 0$; while, when $p < 0$, the solution to \eqref{eq:normal_acceleration} is \emph{indeterminate} or \emph{inconsistent}.
Next, we describe each solution mode in greater detail.
\begin{enumerate}[(i)]
\item \emph{Sliding, $p > 0$, $b < 0$.} --- The end effector is in contact with the belt, i.e.~$z_\R{n} = -H$, $f_\R{n}=-\frac{b}{p}$, and possibly $\dot{z}_\R{t} \ne 0$.
\item \emph{Flight, $p > 0$, $b > 0$.} --- Either $z_\R{n} > 0$, or $z_\R{n} = -H$ and $\ddot{z}_\R{n} > 0$.
\item \emph{Indeterminate, $p < 0$, $b > 0$.} --- The solution is not unique; nonetheless, according to \cite{elkaranshawy2017solving}, when simulating the system, it is possible to resolve the indeterminate mode into a flight mode.
\item \emph{Inconsistent, $p < 0$, $b < 0$.} --- Given the signs of $p$ and $b$, we would have $\ddot{z}_n < 0$, which, recalling that $z_\R{n} = -H$, is not physically feasible, because both the robot and the belt are assumed to be rigid.
This troublesome scenario can be resolved, as explained in \cite{zhao2008experimental}, as an \emph{impact without collision} \cite{GENOT1998}, in which $\dot{z}_\R{n}$ turns from zero to positive, determining the lift-off of the end effector, followed by a succession of bounces on the belt. 
\end{enumerate}
In Figure \ref{fig:phase_plane_open_loop}, we provide an example of the regions associated to each of the four solution modes.


\subsection{Bifurcation diagrams}
\label{subsec:bifurcation_diagrams}

\begin{figure}[t]
    \centering {
        \includegraphics{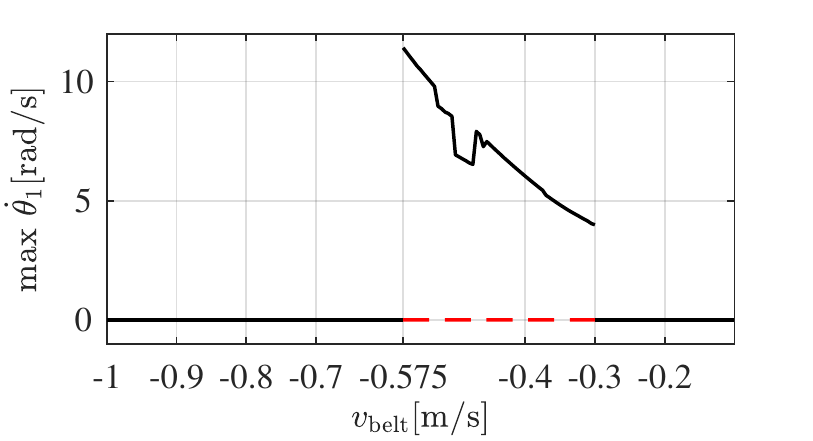}
        \caption{Bifurcation diagram with $\mu = 0.6$. The black solid line corresponds to initial conditions $\B{x}_{0,\R{a}} = [32 \ 0 \ 18.27 \ 0 ]\T$, whereas the dashed red line corresponds to $\B{x}_{0,\R{d}} = [-11.4 \ 0 \ -35.1 \ 0 ]\T$.}
        \label{fig:bifurcation_open_loop}
    }
\end{figure}

\begin{figure}[t]
    \centering {
        \subfigure[]{\label{fig:poincare_map_large}
            \includegraphics{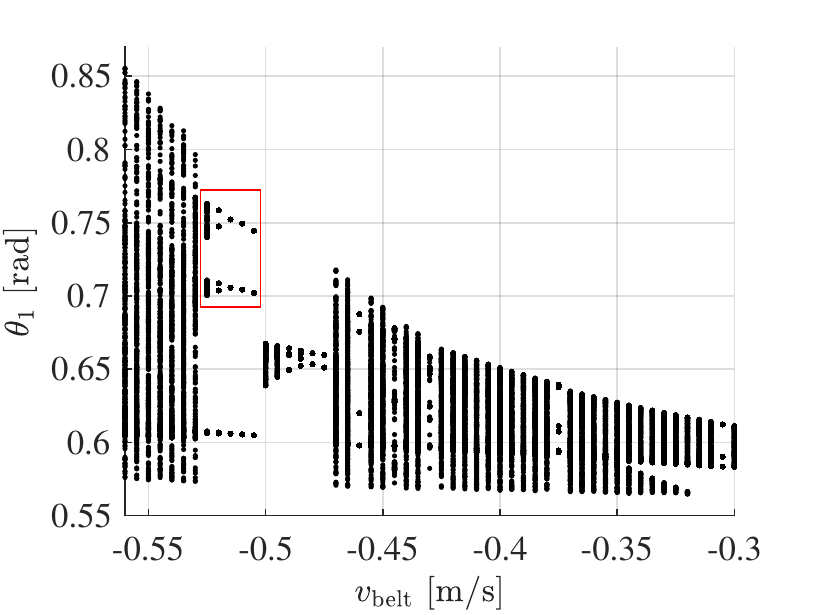}}
        \subfigure[]{\label{fig:poincare_map_detail}
            \includegraphics{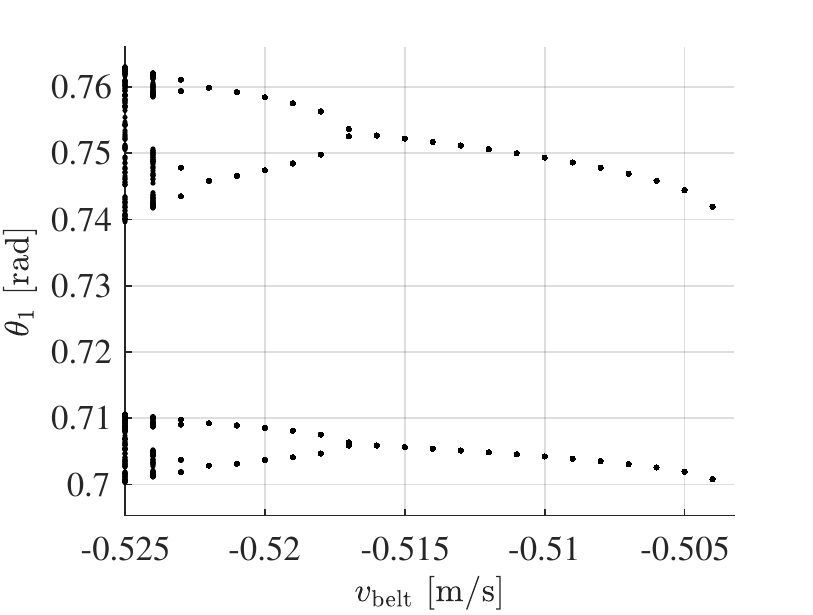}}
        \subfigure[]{\label{fig:poincare_map_chaos_periodicity}
            \includegraphics[scale=0.4]{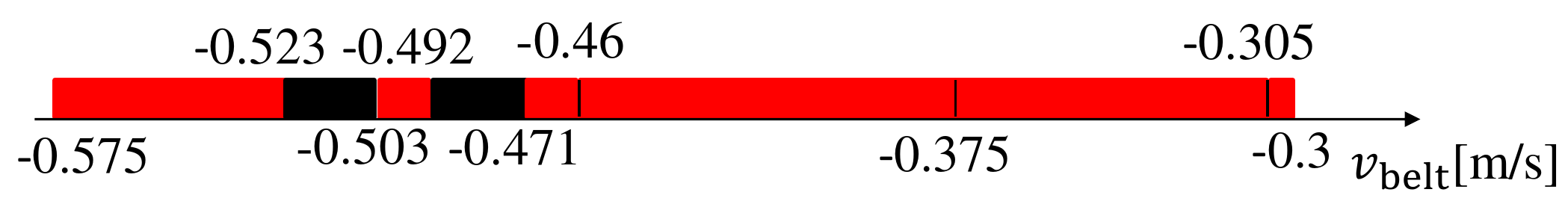}}
    }
    \caption{Bifurcation diagram with $\mu = 0.6$ and initial conditions $\B{x}_{0,\R{a}} = [32 \ 0 \ 18.27 \ 0 ]\T$. (a) is the full picture, while (b) is an enlargment of the portion in the red box traced in (a); (c) depicts the type of the asymptotic behaviour: red represents a chaotic dynamics, whereas black stands for periodic motion.}
    \label{fig:poincare_map}
\end{figure}

To better understand the occurrence of the paradox causing the lift-off of the end effector, we traced a two-dimensional numerical bifurcation diagram in the parameter space consisting of the friction coefficient $\mu$ and the speed of the belt $v_\R{belt}$.

The system was simulated using event-detection routines available in Matlab to detect transitions between each of the solution modes described in Section \ref{subsec:model_description}. 
The bifurcation diagram was constructed via a brute-force method \cite{jacobson2002stability} by simulating the system from a set of random initial conditions for parameters selected in a grid defined by the ranges $0.1\leq \mu \leq 1$ and $-1 \leq v_{\R{belt}} \le -0.1$, with steps of 0.1, and 0.005, respectively.
In each run, the state values are recorded, after a transient time of 250 s, over a time interval of 50 s. 
Then, if in a certain run $\max \dot{\theta}_1 > 0$, it means that the parameter values used in that simulation are such that persistent bouncing motion manifests, which is undesired.

We observed that the bounces appear only for $\mu \geq \mu_\R{c} = 0.4$, that is $\R{max} \dot{\theta}_1 = 0$ if $\mu < \mu_\R{c}$, for all values of $v_\R{belt}$. 
Moreover, we verified that, when $\mu \ge \mu_\R{c}$, features of the bouncing motion such as duration and (a)periodicity depend only on $v_{\R{belt}}$. 
Given that the bifurcation diagram is flat for $\mu < \mu_\R{c}$, and independent of $\mu$ provided that $\mu \ge \mu_\R{c}$, we only present a two-dimensional section of the diagram, in Figure \ref{fig:bifurcation_open_loop}, where $\mu = 0.6$ was considered.
We note that (i) not all initial conditions trigger the bounces (see the red dashed line), and that (ii) bounces are present only when $-0.575 \le v_\R{belt} \le -0.3$.

In order to gain greater knowledge on the specific behaviour of the system when $-0.575 \le v_\R{belt} \le -0.3$ (still with $\mu = 0.6$), we traced a second more detailed bifurcation diagram in Figure \ref{fig:poincare_map}, in which we plot the value of $\theta_1$ when $\dot{\theta}_1$ turns from negative to positive, as a function of the parameter value.
The diagram shows the presence of both periodic and chaotic solutions, providing evidence for the onset of complex seemingly aperiodic behaviour in the parameter regions depicted in red in Figure \ref{fig:poincare_map_chaos_periodicity}.

\section{Control synthesis}
\label{sec:control_synthesis}

Next, we wish to design a controller able to avoid the onset of the bouncing motion due to the paradox and keep the robot moving in contact with the belt.
This in turn requires using a feedback control to guarantee that $p>0$ and $b>0$ at all times in \eqref{eq:normal_acceleration}.
Without loss of generality, we set $v_\R{belt}= - 0.4$, that is a value that allows the occurrence of the paradox.
Firstly, setting $\mu = 0.6$, in Table \ref{tab:admissible_configurations} we determine analytically the values of $z_\R{t}$ such that $p > 0$; we call these \emph{admissible configurations}, given that indeterminate and inconsistent solutions will not appear for such values of $z_\R{t}$.
Secondly, one should determine, among the admissible configurations, those corresponding to $b < 0$; nevertheless, this task is not easy to achieve analytically, because, differently from $p$, $b$ is also a function of $\dot{\B{q}}$ and $\B{u}$.
However, $b < 0$ can be attained using a control scheme that aims at keeping $f_\R{n} > 0$, as it is easy to verify from \eqref{eq:motion_equation_end_effector_n}, when $\ddot{z}_\R{n} = 0$.
We start by using a simpler PID controller, showing that such strategy can keep the end effector in contact with the belt only in a narrow range of the admissible configurations.
Next, we move to a hybrid force/motion control \cite{Siciliano2008} (that allows the regulation of $f_\R{n}$) and demonstrate that this latter approach guarantees avoidance of the lift-off of the end effector in a wider range of the admissible configurations.

\begin{table}[t]
    \caption{Admissible configurations for $\mu=0.6$}
    \label{tab:admissible_configurations}	
    \begin{center}
        \begin{tabular}{@{}ll@{}}
        	\toprule
        	Elbow position                     & Admissible configurations [m]\\ \midrule
        	elbow up ($\theta_2-\theta_1>0$)   & $-0.184 \leq z_\R{t}< 0.183$\\
        	elbow down ($\theta_2-\theta_1<0$) & $-0.184 \leq z_\R{t} < 0.168$\\ \bottomrule
        \end{tabular}
    \end{center}
\end{table}

\subsection{PID strategy}
\label{subsec:pid_strategy}

\begin{figure}[t]
    \centering {
        {\includegraphics{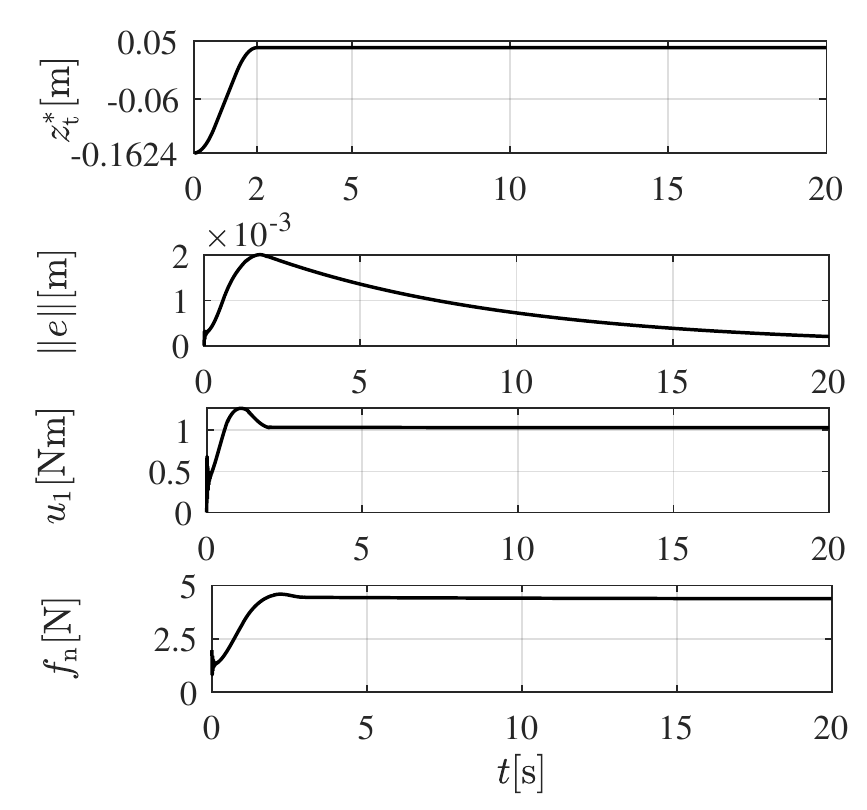}}
        \caption{Simulation with PID control and  $\B{x}(t = 0) = \B{x}_{0,\R{d}}$.}
        \label{fig:pid}
    }
\end{figure}

\begin{figure}[t]
    \centering
    \includegraphics{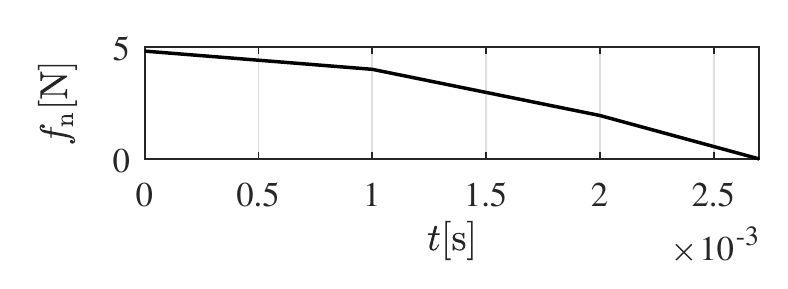}
    \caption{Simulation with PID control, $\B{x}(t = 0) = \B{x}_{0,\R{u}}$ and $z_\R{t}^*$ as in Figure \ref{fig:pid}.}
    \label{fig:failure}
\end{figure}

For the sake of simplicity, we started by considering a simpler PID control approach to test its feasibility to solve the control goal.
Let $\B{z}^*$ be the reference value for the end effector coordinates, $e \triangleq z_\R{t}^*-z_\R{t}$ a reference error, and $\B{q}' \triangleq \begin{bmatrix} \theta_1 & \theta_2 - \theta_1 \end{bmatrix}\T$.
Say $\B{q}'^*$ the reference value for $\B{q}'$, computed from $\B{z}^*$
using inverse kynematics as explained in \cite{Siciliano2008}.
Hence, the control terms $u_i$, $i = 1,2$, obtained using a PID control scheme are given by
\begin{equation}
u_i = K_{\R{P},i} (q'^*_i - q'_i) + K_{\R{I},i} \int_{0}^{\tau} (q'^*_i - q'_i) \R{d}t + K_{\R{D},i} (\dot{q}'^*_i - \dot{q}'_i),
\end{equation}
where $K_{\R{P},i}$, $K_{\R{I},i}$, $K_{\R{D},i}$, $i = 1,2$, are constants.
The PID gains were selected heuristically by running a series of numerical simulations from two sets of initial conditions.
These are $\B{x}_{0,\R{d}} = \begin{bmatrix} -11.4 & 0 & -35.1 & 0 \end{bmatrix}\T$ and $\B{x}_{0,\R{u}} = \begin{bmatrix} -35.1 & 0 & -11.4 & 0 \end{bmatrix}\T$, both corresponding to $\B{z} = \begin{bmatrix} -0.1624 & 0 \end{bmatrix}\T$, with the only difference that $\B{x}_{0,\R{d}}$ is an ``elbow down'' posture ($\theta_2-\theta_1<0$) and $\B{x}_{0,\R{u}}$ is an ``elbow up'' posture ($\theta_2-\theta_1>0$).
The gains were adjusted in a trial-and-error process with the aim of obtaining a large value of $Z_{\R{t},\R{sliding}}$, that is the largest value of $z_\R{t}$ such that no lift-off occurs. 
We observed different results, depending on the initial condition. For $\B{x}(t = 0) = \B{x}_{0,\R{d}}$, $Z_{\R{t},\R{sliding}} = 0.0375$, and acceptable values of the gains were found to be $K_{\R{P},1} = 200, K_{\R{I},1} = 25, K_{\R{D},1} = 2$, and $K_{\R{P},2} = K_{\R{I},2} = K_{\R{D},2} = 0$.
The corresponding simulation graphs, are shown in Figure \ref{fig:pid}.
 Differently, for $\B{x}(t = 0) = \B{x}_{0,\R{u}}$, the simulations results showed that the PID control is not able to effectively avoid the onset of the paradox.
As a matter of fact, we could not find values of the control gains such that lift-off was avoided.
An example is visible in Figure \ref{fig:failure}, where the time evolution of the normal reaction $f_\R{n}$ is depicted; notice that it eventually becomes zero, meaning that the end effector detaches from the belt.



\subsection{Hybrid force/motion control}
\label{subsec:hybrid_motion_force_control}

\begin{figure}[t]
    \centering
    \includegraphics[scale=0.35]{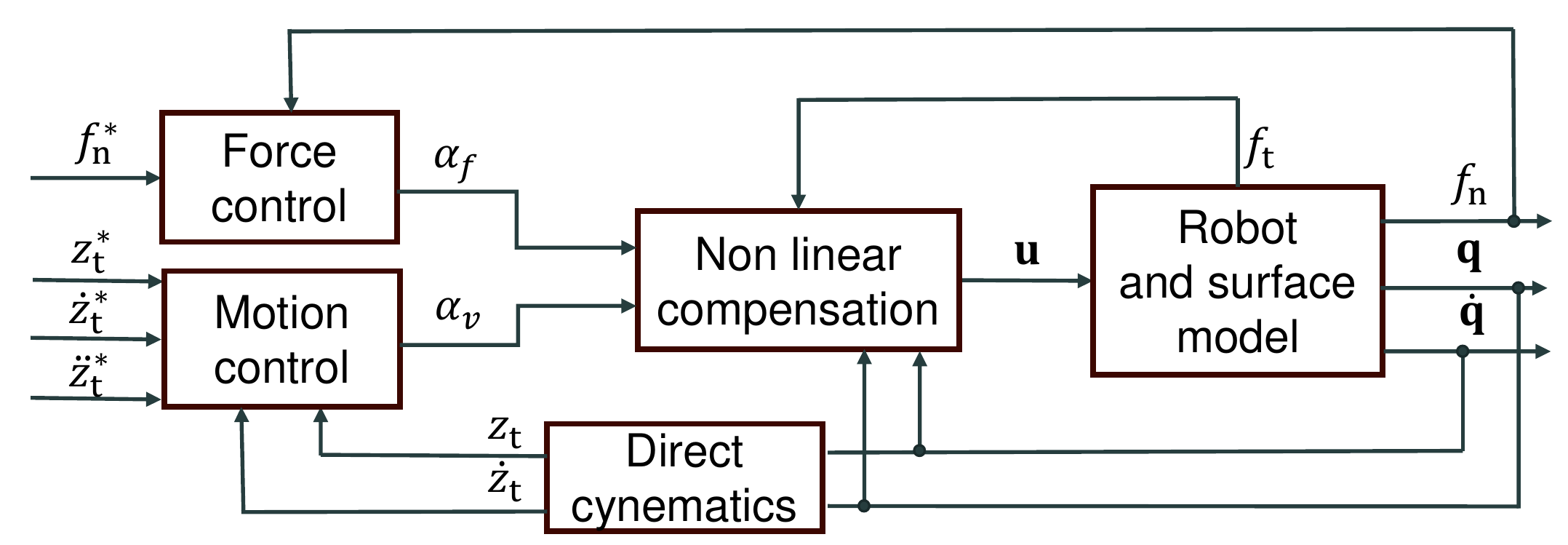}
    \caption{Hybrid force/motion control scheme.}
    \label{fig:force_motion_control_scheme}
\end{figure}

Next, we show that better performance can be achieved with a force/motion control scheme \cite{Siciliano2008}, since it allows to regulate the value of the normal reaction $f_\R{n}$ in addition to the end effector's tangential position $z_\R{t}$.
In particular, defining the unit vectors $\hat{\B{i}}_x \triangleq \begin{bmatrix} 1 & 0 \end{bmatrix}\T$ and $\hat{\B{i}}_y \triangleq \begin{bmatrix} 0 & 1 \end{bmatrix}\T$, associated to the $x$ and $y$ Cartesian axes, the control action is given by 
\begin{equation} \label{eq:force_motion_control}
\B{u} = \B{w} + \B{c} + \B{M} \B{J}^{-1} \left( - \dot{\B{J}} \dot{\B{q}} + \hat{\B{i}}_x \alpha_v \right) + \B{J}\T \left( - \hat{\B{i}}_x f_{\R{t}} - \hat{\B{i}}_y \alpha_{f} \right).
\end{equation}
Note that, in \eqref{eq:force_motion_control}, on the right-hand side, the first, second, and fifth terms compensate corresponding terms in \eqref{eq:motion_equation_end_effector}.
Differently, the fourth and and sixth terms are used to assign dynamics for $z_\R{t}$ and $f_\R{n}$, respectively. Specifically, letting $f_\R{n}^*$ be a reference value for the normal reaction, we choose
\begin{align}
\alpha_v =& \ddot{z}_\R{t}^* + K'_{\R{P}} (z^*_{\R{t}} - z_\R{t}) + K'_\R{D} (\dot{z}^*_\R{t} - \dot{z}_\R{t}), \\
\alpha_{f} =& f^*_\R{n} + K'_\R{I} \int_{0}^{\tau} (f^*_\R{n} - f_\R{n}) \R{d}t,
\end{align}
where $K'_\R{P}$, $K'_\R{D}$, and $K'_\R{I}$ are control gains.
A block diagram of the hybrid force/motion control scheme is illustrated in Figure \ref{fig:force_motion_control_scheme}.

\begin{figure}[t]
    \centering
    \includegraphics{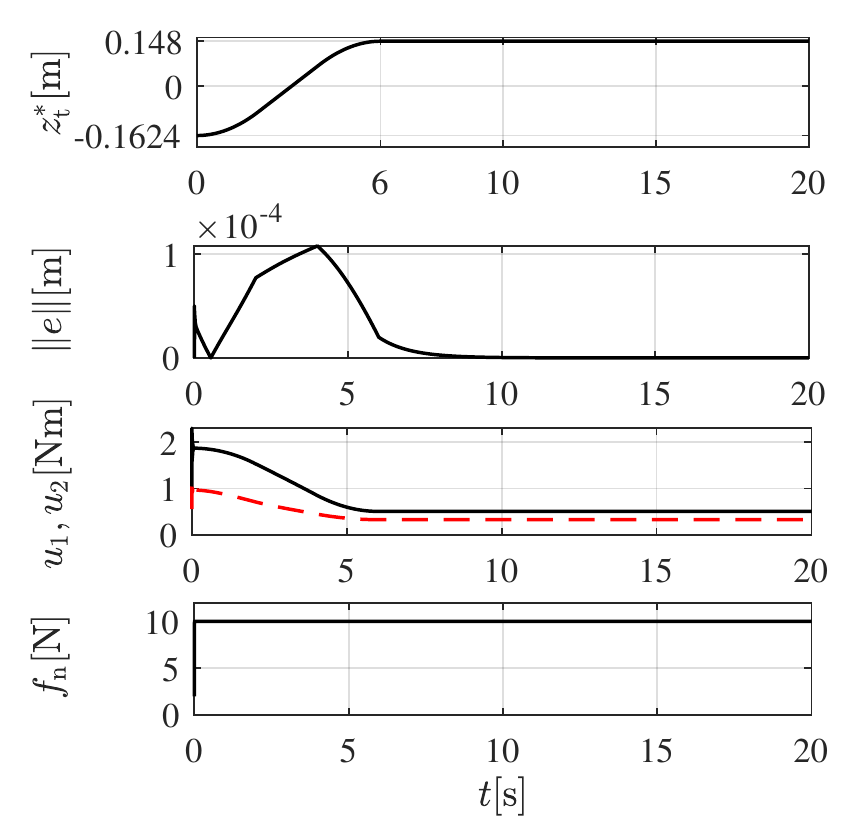}
    \caption{Simulation with force/motion control, $\B{x}(t = 0) = \B{x}_{0,\R{d}}$ and $f_\mathrm{n}^*=10$ N.
    In the third panel from the top, the black solid line is $u_1$, whereas the red dashed line is $u_2$.}
    \label{fig:hybrid}
\end{figure}

\begin{figure}[t]
    \centering
    \includegraphics{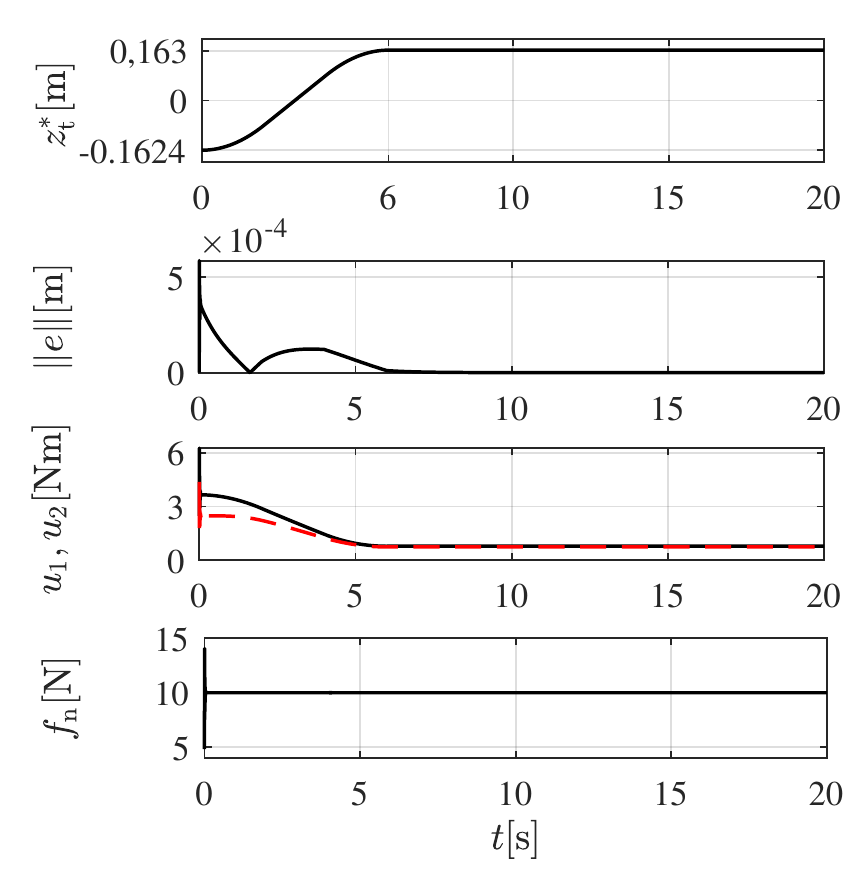}
    \caption{Simulation with force/motion control, $\B{x}(t = 0) = \B{x}_{0,\R{u}}$ and $f_\mathrm{n}^*=10$ N.
    In the third panel from the top, the black solid line is $u_1$, whereas the red dashed line is $u_2$.}
    \label{fig:hybridone}
\end{figure}

\begin{figure}[t]
    \centering
    \includegraphics{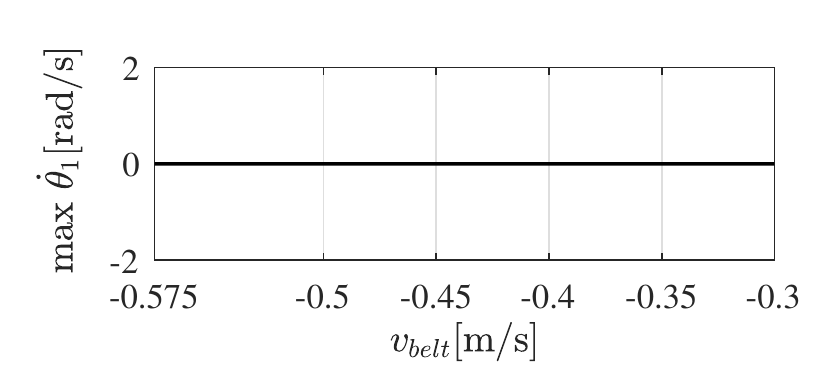}
    \caption{Closed loop bifurcation diagram obtained with $\B{x}(t = 0) = \B{x}_{0,\R{d}}$, the same references as that in Figure \ref{fig:hybrid}, and in the presence of the force/motion control.}
    \label{fig:bifurcation_closed_loop}
\end{figure}

To test the performance of the control system, we ran a series of simulations from the same initial conditions used to validate the PID control strategy; the control gains being selected heuristically as  $K'_\R{P} = K'_\R{D} = 900, K'_\R{I} = 650$ for $\B{x}(t=0)=\B{x}_{0,\R{d}}$ and $K'_\R{P} = K'_\R{D} = 900, K'_\R{I} = 0$ for $\B{x}(t=0)=\B{x}_{0,\R{u}}$.
The numerical results showed that, for $\B{x}_{0,\R{d}}$, $Z_{\R{t},\R{sliding}} = 0.148$, whereas, for $\B{x}_{0,\R{u}}$, $Z_{\R{t},\R{sliding}} = 0.163$, which are both higher than the values obtained with the PID, meaning that the force/motion control scheme allows the robot to operate in a wider range of configurations.
Examples of simulations are shown in Figures \ref{fig:hybrid} and \ref{fig:hybridone}, representing the results of the simulations starting from $\B{x}_{0,\R{d}}$ and $\B{x}_{0,\R{u}}$, respectively.
Moreover, we verified that when using the present control strategy, the persistent bouncing motion is suppressed for all $v_\R{belt} \in [-1, -0.1]$.
This is shown in the closed-loop bifurcation diagram in Figure \ref{fig:bifurcation_closed_loop}, which can be compared with that in Figure \ref{fig:bifurcation_open_loop}, representing the bifurcation diagram for the open loop system.
As expected, the closed-loop system remains in contact with the belt over the entire parameter region of interest without any bifurcation to persistent bouncing motion. 


\section{Conclusions}
\label{sec:conclusions}
We dealt with the analysis and control of the Painlev\'{e} paradox in a two-links robot in contact with a moving belt.
The paradox determines occasional lift-off of the tip of the robot, which is undesired for a number of applications, like cutting or objects moving. 
We started by conducting a bifurcation study varying the belt speed, finding that some values determine a chaotic motion of the end effector, while for others the motion is a periodic bouncing.
Then, we used the results of the bifurcation analysis to inform the control design and proposed two control schemes, a PID controller and a hybrid force/motion control strategy, which we compared through numerical simulations.
We showed that the latter strategy is effective in preventing the paradox from occurring and hence guaranteeing that end effector of the robot stays in contact with the belt over a wider parameter range with respect to the PID.

\bibliographystyle{IEEEtran}
\bibliography{references_painleve_ecc2019}

\end{document}